\documentclass[final,1p,times]{elsarticle} 

\usepackage{graphicx}
\usepackage{amssymb} 
\usepackage{amsthm} 
\usepackage{lineno}
\usepackage{subfigure}
\usepackage{bm}

\journal{Nuclear Physics A} 

\begin{document}

\begin{frontmatter} 
  \title{Imprinting Quantum Fluctuations on Hydrodynamic Initial Conditions}
  \author[OSU,Duke]{J.S. Moreland} 
  \author[OSU]{Z. Qiu}
  \author[OSU]{U. Heinz}
  \address[OSU]{The Ohio State University, 191 West Woodruff Avenue, Columbus OH 43210, USA}
  \address[Duke]{Duke University, Physics Bldg., Science Dr., Box 90305, Durham, NC 27708, USA}

\begin{abstract}
We have developed an algorithm to imprint quantum fluctuations onto the initial transverse energy density profile according to a given two-point covariance function. 
Using as an example MC-KLN initial conditions  with added fluctuations satisfying the covariance function derived in \cite{MS}, we find that effects from sub-nucleonic 
gluon field fluctuations on the eccentricity harmonics $\epsilon_{n}$ vary strongly with the gluonic correlation length controlled by the saturation momentum $Q_{s}$. 
Varying $Q_{s}$ over the range probed in Au+Au collisions at RHIC, we find gluon fluctuation induced enhancements of the eccentricity coefficients ranging from 
10 to 20\% in central collisions.
\end{abstract}

\end{frontmatter}

\section{Event-by-event fluctuations}

The importance of event-by-event fluctuations in the initial conditions context of heavy-ion collisions was first pointed out by Miller and Snellings who added nucleon position fluctuations to existing smooth (i.e. ensemble-averaged) Glauber initial conditions \cite{MillerSnellings, FirstGlb}. These fluctuations explain the experimental observation of non-vanishing anisotropic flow in central Cu-Cu and Au-Au collisions \cite{vn}
and of odd flow harmonics $\{v_3, v_5,...\}$ \cite{Alver:2010gr}.

Recently attention has turned to a new source of event-by-event fluctuations, namely fluctuations in the transverse distribution of color charge within the colliding nucleons. These fluctuations are evidenced by  large multiplicity fluctuations observed in minimum bias $pp$ collisions and suggest that sub-nucleonic degrees of freedom may also play an important role in determining the event-by-event geometry of the initial state in nucleus-nucleus collisions \cite{Tribedy:2010ab,Schenke:2012wb,KNO}.

\section{From nucleonic to sub-nucleonic fluctuations}

In this work, we develop a toy model for imprinting sub-nucleonic fluctuations on the transverse energy density profiles produced in relativistic heavy-ion collisions. The study is motivated by recent work of
M\"uller and Sch\"afer in which they calculate the mean normalized covariance function $\mbox{Cov}[\epsilon(r)/\epsilon_0]$ for the transverse energy density fluctuations of gluon fields in central Au-Au collisions at 200 GeV \cite{MS}. The authors approximate the collision system by two infinite slabs of nuclear matter with fixed gluon saturation momentum $Q_{s}$ for which they take the value corresponding to the nuclear thickness function in the center of a central Au-Au collision.

We texture a given transverse energy density profile with additional fluctuations using a Turning Band Gaussian random field simulator (TBSIM) \cite{TBSIM} that includes several configurable covariance functions. The M\"uller-Sch\"afer covariance $\mbox{Cov}[\epsilon(r)/\epsilon_0] = (\Delta \epsilon(r)/ \epsilon_0)^2$ is well described by the Cauchy covariance included in TBSIM,
\begin{equation}
 \mbox{Cov}[r]=C(1+(r/a)^2)^{-b}
\end{equation}
with fit paramaters $C=0.4679$, $a=0.2878$ fm and $b=1.7732$. 

\begin{figure}[h!]
\begin{center}
\includegraphics[width=56mm]{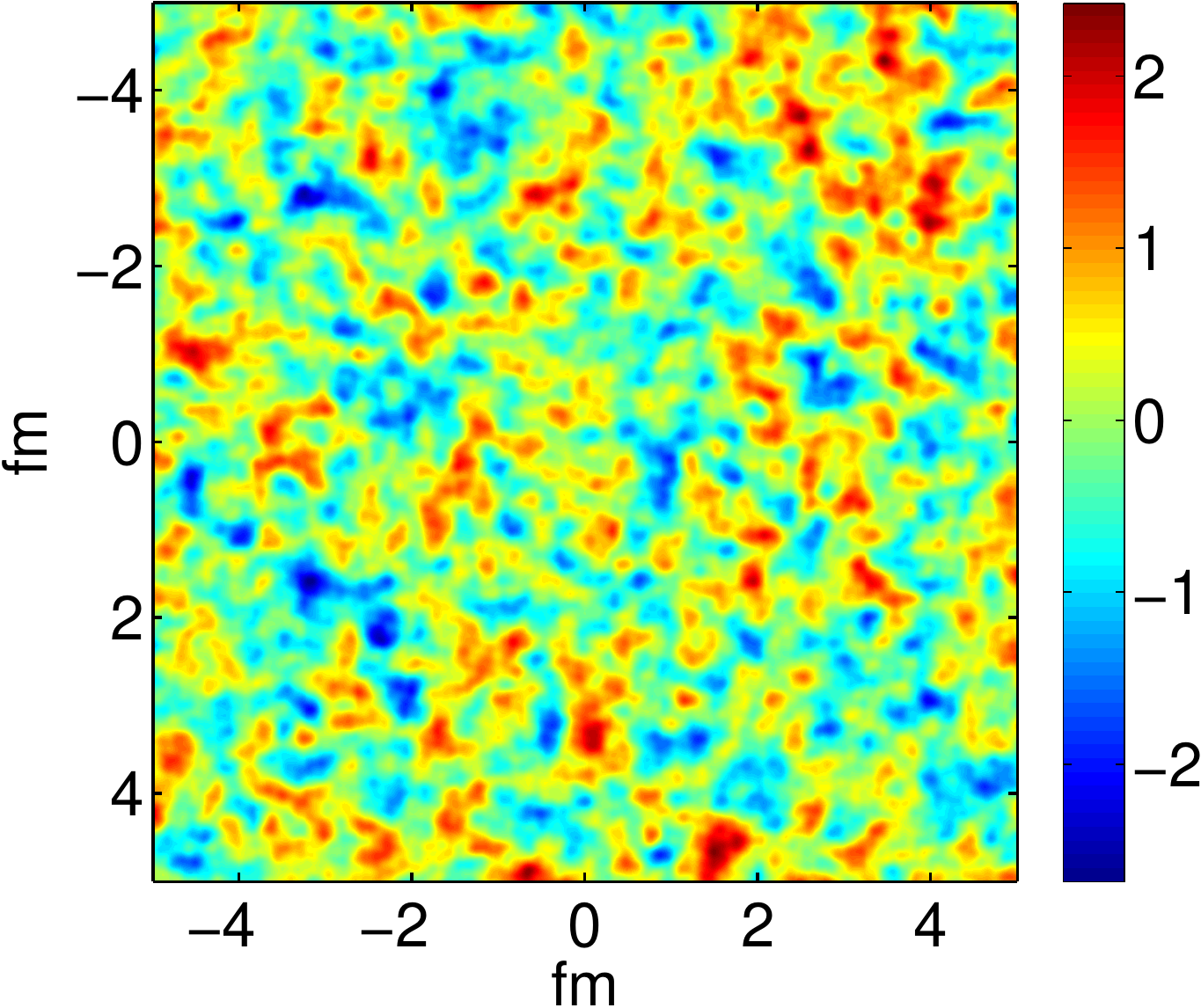}
\hspace{0.1 in}
\includegraphics[width=66mm, height=48mm]{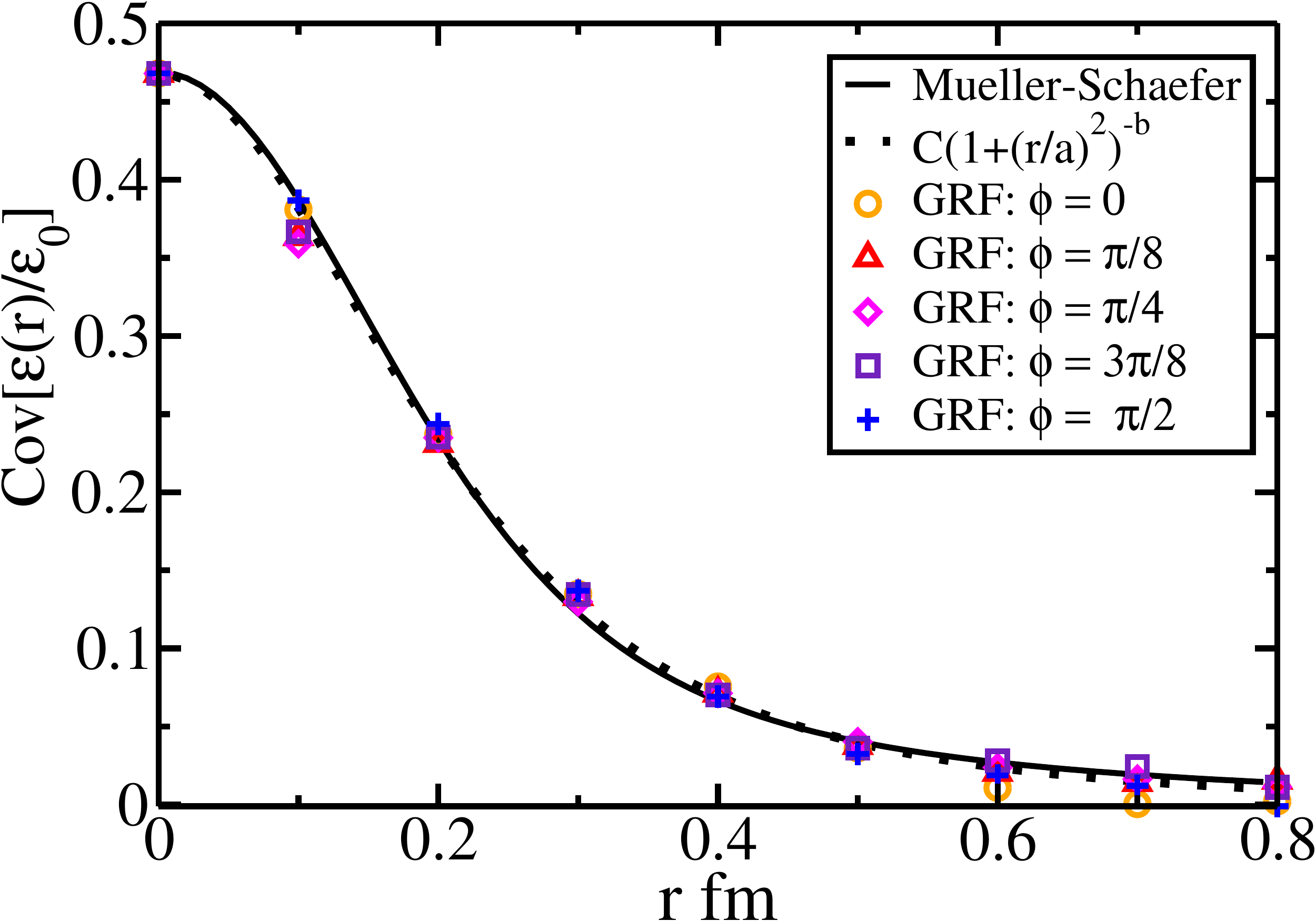}
\caption{(Color online) {\sl Left:}
10\,fm${\times}$10\,fm section of the TBSIM Gaussian Random Field for the normalized energy density 
$\epsilon(\bm{r})/\epsilon_0{-}1$ with zero mean and a Cauchy covariance fitted to the M\"uller-Sch\"afer covariance.  {\sl Right:} Two-point covariance of the GRF as a function of distance $r$ at various azimuthal angles $\phi$ (symbols), compared with the M\"uller-Sch\"afer function (solid) and its Cauchy Covariance fit (dotted).
\label{F1}
}
\end{center}
\end{figure}

Using TBSIM we generate a large Gaussian random field (GRF) with the desired covariance on a $4000 \times 4000$ lattice with grid spacing $\Delta x{\,=\,}0.1$\,fm. The left panel in Fig.~\ref{F1} shows a small slice of this GRF; in the right panel its two-point correlation function is compared to the M\"uller-Sch\"afer covariance and its Cauchy fit.
 
%
 
\begin{figure}[h!]
\begin{center}
\includegraphics[width=56mm]{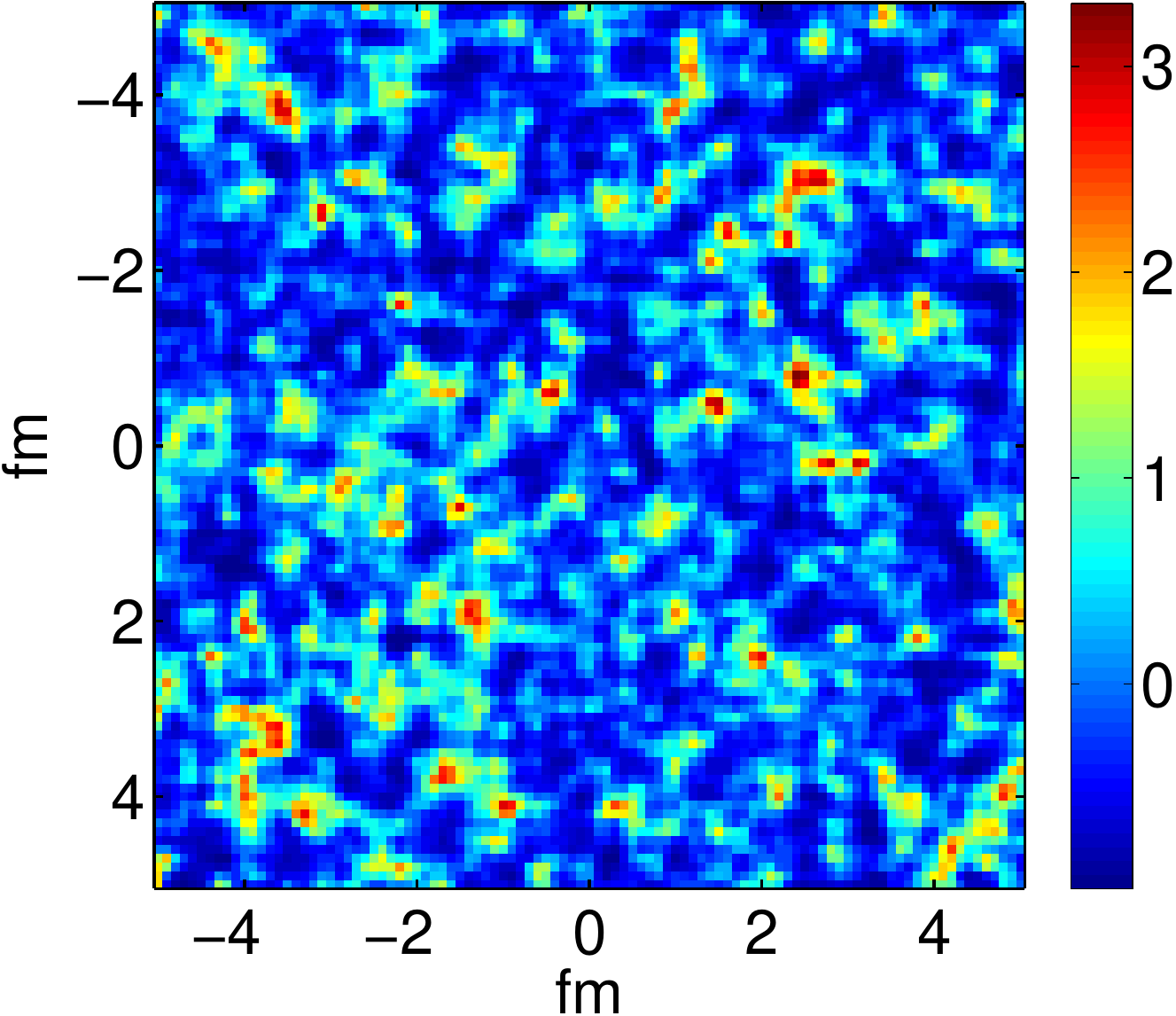}
\hspace{0.1 in}
\includegraphics[width=66mm, height=49mm]{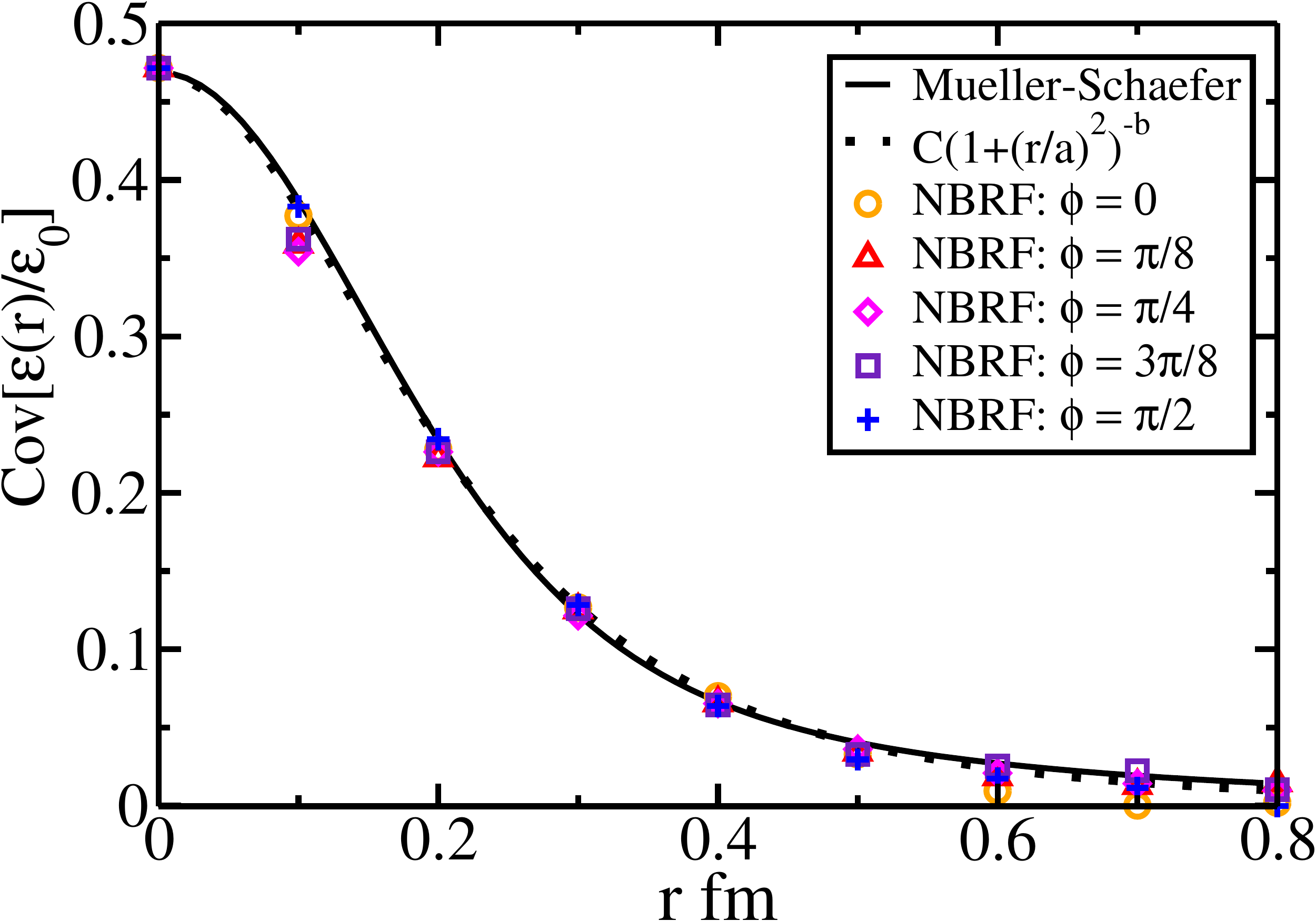}
\caption{(Color online) 
Same as Fig.~\ref{F1}, but for the NBD random field. Note that in the left panel the plotted quantity  $\epsilon(\bm{r})/\epsilon_0{-}1$ is now ${\geq}-1$ everywhere.
\label{F2}
}
\end{center}
\end{figure}

Due to the tail of the Gaussian distribution, there is a $7.2\%$ chance that fluctuations about the mean fall into the unphysical region of negative energy density. These unphysical fluctuations can be eliminated by mapping the Gaussian random variable onto a positive definite negative binomial distribution (NBD). There is a long history of modeling fluctuations in $pp$ collisions with negative binomial distributions (for recent examples see \cite{Tribedy:2010ab,KNO}). Writing 
\begin{equation}
 \mathrm{NBD}(\bar{n},k;n) = \frac{\Gamma(k+n)}{\Gamma(k)\Gamma(n+1)} \frac{\bar{n}^n k^k}{(\bar{n}+k)^{n+k}} ,
\end{equation}
where $n$ is the sampled value, $\bar{n}$ is its mean and $k$ controls its variance, we identify $n/\bar{n}$ with $\epsilon(r)/\epsilon_0$. In the limit of large $\bar{n}$ (we take $\bar{n}{\,=\,}100$), the NBD becomes a continuous function $P_\mathrm{NBD}(y)$ of the reduced variable $y{\,=\,}n/\bar{n}$ whose width parameter $k$ we adjust such that its variance $\bigl\langle\bigl(\frac{n}{\bar{n}}{-}1\bigr)^2\bigr\rangle{\,=\,}\frac{1}{k}{+}\frac{1}{\bar{n}}$ agrees with the squared Gaussian width $\mathrm{Cov}[\epsilon(r)/\epsilon_0]|_{r=0}{\,=\,}(\Delta\epsilon(0)/\epsilon_0)^2{\,=\,}(0.684)^2$. The mapping is now achieved by replacing each value of $x{\,=\,}\epsilon/\epsilon_0$ from the Gaussian random field by a new value $y{\,=\,}\epsilon_\mathrm{new}/\epsilon_0$ such that the cumulative Gaussian distribution function at $x$ coincides with the cumulative NBD distribution at $n/\bar{n}{\,=\,}\epsilon_\mathrm{new}/\epsilon_0{\,=\,}y$: 
\begin{equation}
 \int_{-\infty}^{x} P_\mathrm{Gauss}(x')\, dx' = \int_0^{y} P_\mathrm{NBD}(y')\, dy'.
\end{equation}
The resulting negative binomial random field $\epsilon(\bm{r})/\epsilon_0$ is positive definite and retains the two-point covariance embedded in the original GRF. A small section of this field and its two-point correlation function are shown in Fig.~\ref{F2}.
%
%

To imprint these energy density fluctuations onto the initial conditions for a heavy-ion collision, we take the transverse energy densities $dE/d^2r_\perp dy$ for events generated with the MC-KLN model \cite{MCKLN}
and multiply them with the NBD random $\epsilon(\bm{r})/\epsilon_0$ field taken from randomly selected and appropriately sized sections of the final $4000{\,\times\,}4000$ field grid: 
\begin{equation}
  \frac{dE_{fluct.}(\bm{r})}{d^2r\, dy} = 
  \frac{dE_{KLN}(\bm{r})}{d^2r\, dy} \times \frac{\epsilon(\bm{r})}{\epsilon_0}.
\end{equation}

\section{Results and Conclusions}

\begin{figure}[h!]
 \begin{center}
 \includegraphics[width=60mm]{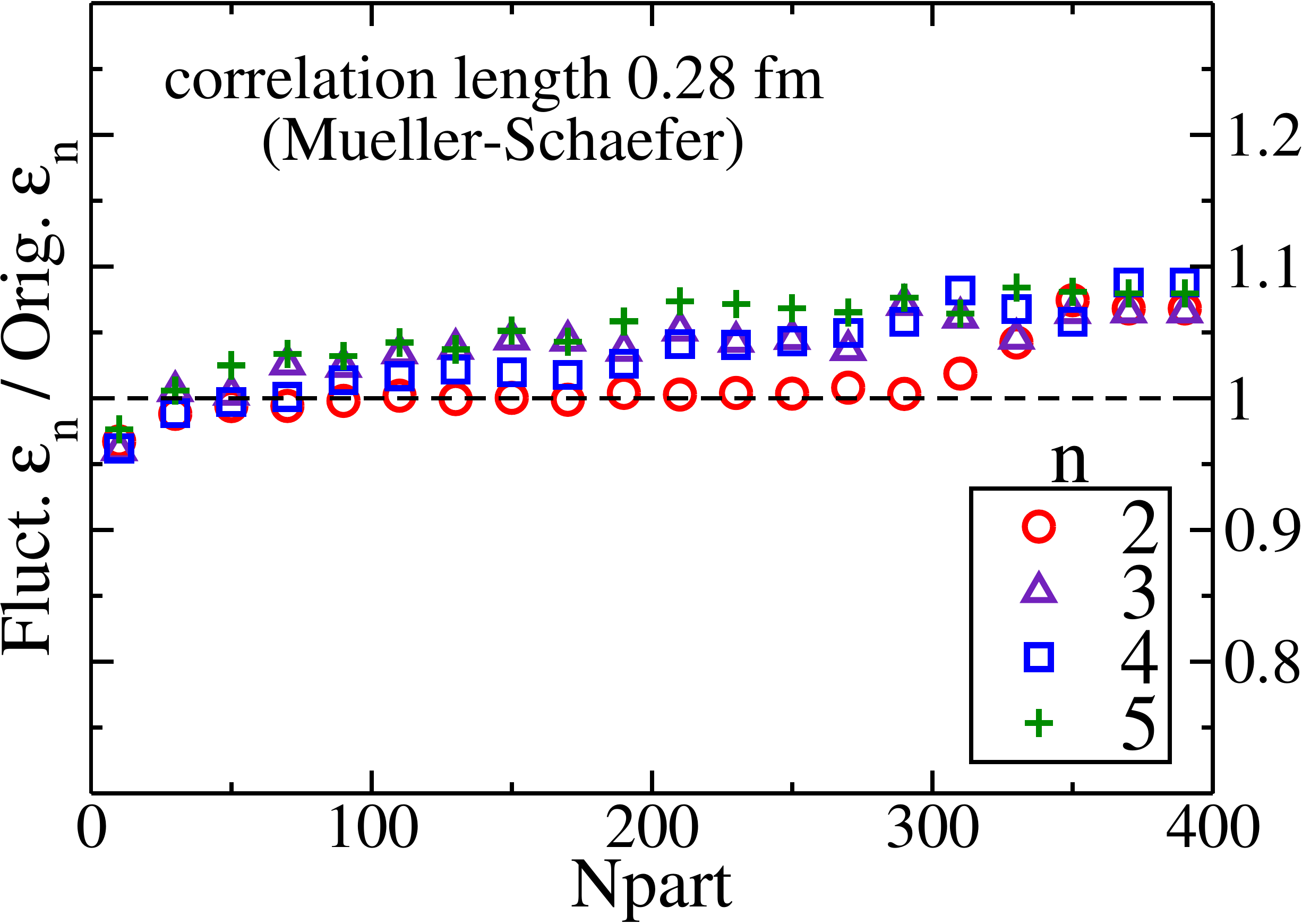}
 \hspace{0.1 in}
 \includegraphics[width=60mm]{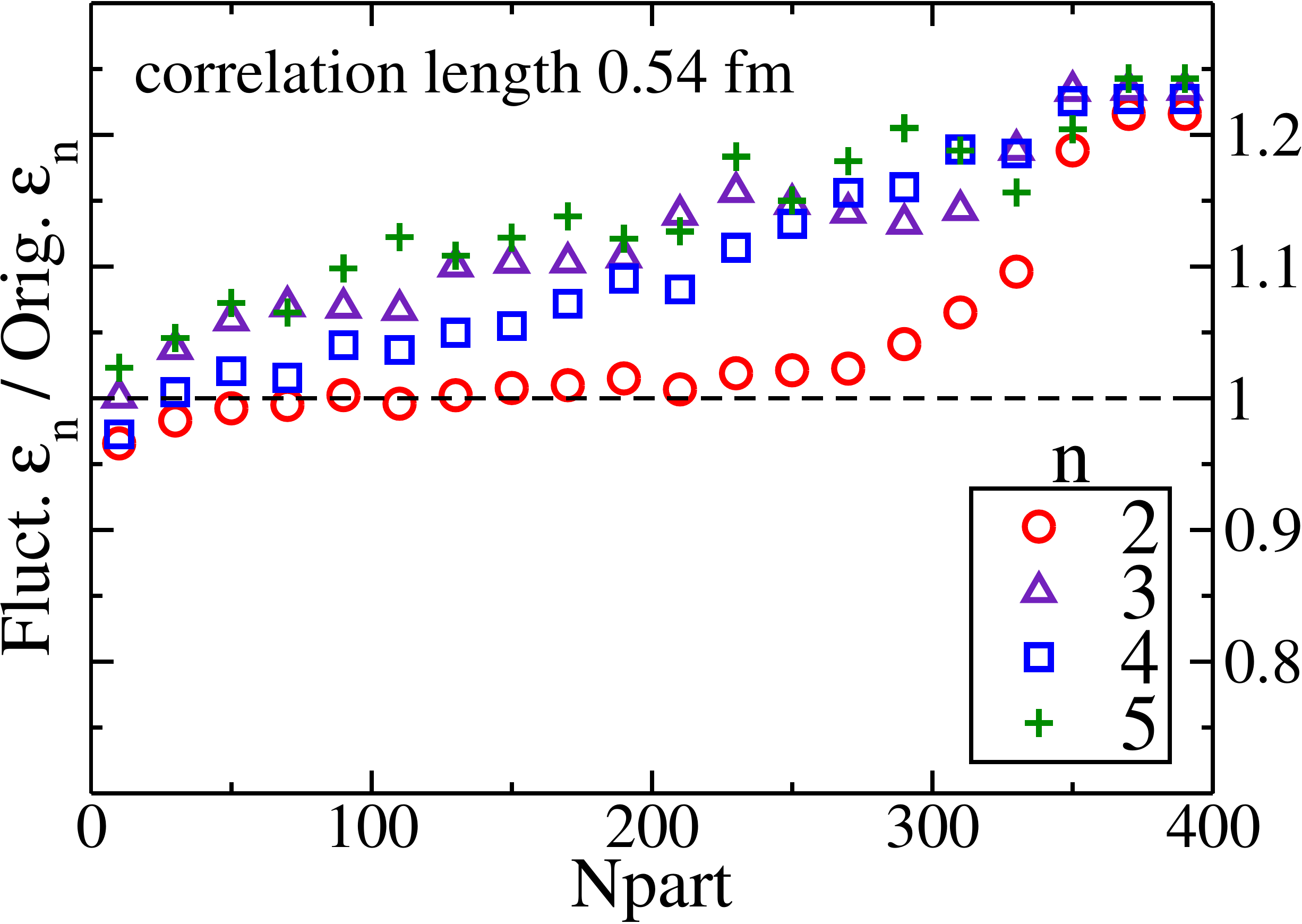}
 \caption{(Color online) Centrality dependence of the gluon-field fluctuation induced enhancement
     of $\epsilon_n$ in $200\,A$\,GeV Au-Au events, for a correlation length $a{\,=\,}0.28$\,fm as 
     used in \cite{MS} (left panel) and a value $a{\,=\,}0.54$\,fm as large as the nucleon size (right panel).
 \label{F3}
 }
 \end{center}
 \end{figure}

This texturing procedure was applied to $20,000$ MC-KLN Au-Au events at $\sqrt{s}{\,=\,}200\,A$\,GeV
(using the MC-KLN code with Gaussian nucleons of width $\sigma = 0.54$\,fm), partitioned into equally sized bins in the number of participants of width $\Delta N_\mathrm{part}{\,=\,}1000$. For both the textured and untextured events we compute the harmonic eccentricity coefficients $\epsilon_n$ using the definition
\begin{equation}
 \epsilon_{n} e^{i n \Phi_n}= - \frac{\int r\,dr\,d\phi\,r^2\,e^{i n \phi}\,\frac{dE(r,\phi)}{d^2r \,dy}}
                                                       {\int r\,dr\,d\phi\,r^2\,\frac{dE(r,\phi)}{d^2r\, dy}}
\end{equation}
and average them over the event ensemble. The ratios of these averages are plotted in Fig.~\ref{F3} for the harmonics $n{\,=\,}2,\dots,5$ as functions of $N_\mathrm{part}$.

%

For the nominal correlation length $a{\,=\,}0.28$\,fm , shown in the left panel, we see that the gluon-field fluctuations induce only a small increase in the eccentricity harmonics $\epsilon_n$, reaching 5-10\% in central collisions and falling off in more peripheral ones. We should note, however, that our fluctuation texture assumes constant (i.e. position-independent) $Q_s$ or $a$, with a value expected (on average) in the center of central Au-Au collisions. More realistically, $a^2$ should vary inversely with the position-dependent nuclear thickness function (which, on average, is largest in the fireball center):
\begin{equation}
 a^2 \propto 1/Q_s^2 \propto 1/T(\bm{r}).
\end{equation}
Consequently, the M\"uller-Sch\"afer texture with constant $a{\,=\,}0.28$\,fm provides an approximate \emph{lower} bound on the eccentricity enhancement caused by 
sub-nucleonic fluctuations in $200\,A$\,GeV Au-Au collisions. 

To obtain an estimated upper bound on the eccentricity enhancement caused by sub-nucleonic fluctuations, we inflate the correlation length $a$ to the radius of our 
Gaussian nucleons, $\sigma{\,=\,}0.54$\,fm and repeat the texturing procedure. (This amounts to reducing $Q_s$ by about a factor 2.) As seen in the right panel of 
Fig.~\ref{F3}, with the larger correlation length the gluon fluctuations increase the eccentricities $\epsilon_n$ by larger factors, reaching now 20-25\% in central 
collisions (and again falling off in peripheral ones). For both values of the correlation length, the sub-nucleonic fluctuation effects on $\epsilon_n$ are strongest 
in central collisions; in more peripheral collisions, fluctuations in the nucleon positions dominate the fluctuation effects on $\epsilon_n$.
 
To summarize, we generated a toy model to analyze the effects of sub-nucleonic color fluctuations on the centrality dependent eccentricty harmonics $\epsilon_n$ for MC-KLN initial conditions. While we qualitatively confirm earlier findings \cite{Schenke:2012wb,KNO} that such fluctuations tend to increase the $\epsilon_n$, only relatively small enhancements (smaller than those reported in \cite{KNO}) are found for realistic values of the gluon field correlation length. Correlations over larger distances generate larger eccentricities. Due to the assumption of a position-independent saturation momentum $Q_s$ or correlation length $a$, our implementation of sub-nucleonic fluctuations is much less realistic than the one in the IP-Glasma model of Refs.~\cite{Tribedy:2010ab,Schenke:2012wb}; it has, however, the advantage of allowing us to turn the sub-nucleonic fluctuations on and off at will and thus to study their effects on $\epsilon_n$ in isolation.  

{\bf Acknowledgements:} This work was supported by the U.S. Department of Energy under Grants No. \rm{DE-SC0004286} and (within the framework of the JET Collaboration) \rm{DE-SC0004104}, and by the Ohio Supercomputer Center.

\end{document}